# Using Large Language Models to Assign Partial Credit to Students' Explanations of Problem-Solving Process: Grade at Human Level Accuracy with Grading Confidence Index and Personalized Student-facing Feedback

Zhongzhou Chen* and Tong Wan

*Department of Physics, University of Central Florida, Orlando, Florida 32816, USA*

**ABSTRACT**.
This study examines the feasibility and potential advantages of using large language models, in particular GPT-4o, to perform partial credit grading of large numbers of student written responses to introductory level physics problems. Students were instructed to write down verbal explanations of their reasoning process when solving one conceptual and two numerical calculation problems on in class exams. The explanations were then graded according to a 3-item rubric with each item grades as binary (1 or 0). We first demonstrate that machine grading using GPT-4o with no examples nor reference answer can reliably agree with human graders on 70%-80% of all cases, which is equal to or higher than the level at which two human graders agree with each other. Two methods are essential for achieving this level of accuracy: 1. Adding explanation language to each rubric item that targets the errors of initial machine grading. 2. Running the grading process five times and taking the most frequent outcome. Next, we show that the variation in outcomes across five machine grading attempts as measured by the Normalized Shannon Information Entropy can serve as a grading confidence index. By setting the threshold entropy to 0.4, a human instructor can identify ~40% of all potentially incorrect gradings by reviewing just 10 - 15% of all responses with entropy higher than the threshold. Finally, we show that it is straightforward to use GPT-4o to write clear and detailed explanations of the partial credit grading outcomes. Those explanations can be used as feedback for students, which will allow students to understand their grades and raise different opinions when necessary. Almost all feedback messages generated were rated 3 or above on a 5-point scale by two instructors who had taught the course multiple times. The entire grading and feedback generating process cost roughly $5 per 100 student answers, which shows immense promise for automating labor-intensive grading process by a combination of machine grading with human input and supervision.

## I. INTRODUCTION

The rapid recent advancement in both capability and availability of Generative Artificial Intelligence (GenAI), in particularly large language models (LLM) such as GPT-4, has spurred a growing body of research exploring their applications across many different areas of education [1,2]. One of those areas that has gained a substantial amount of recent interest is Automated Short Answer Grading (ASAG) [3,4] and Automated Long Answer Grading (ALAG) [5]. Both ASAG and ALAG utilize LLM's ability to process and generate natural language to grade large numbers of student written responses, especially for problems in STEM domains including physics. Using GenAI to assist the assigning of students written response has the clear benefit of relieving instructors from time consuming and largely repetitive work of grading [6,7], and could potentially provide students with more detailed and timely feedback on their understanding, which is highly beneficial for learning [8].

Most ASAG or ALAG studies published prior to 2023 used smaller language models that need to first be trained on a corpus of domain specific data (for example see [3,9]) to achieve satisfactory performance. Since 2023, many studies have shifted towards using general purpose language models, also called *foundational models* [10], with the most popular foundation models being GPT-3.5 and GPT-4 [11]. The most significant advantage of using foundational LLMs is that they require no pre-training on existing data, which greatly simplifies the grading process. In most cases, grading one student response simply involves sending a piece of natural language text, called a *prompt*, to the LLM. A prompt usually consists of a problem body, a grading rubric and grading requirements, and a student response. Sometimes one example response-grading pair or more could also be included in the prompt.

It is important to clarify that the term GenAI refers to a family of artificial intelligence models that generates contents ranging from text, computer code, images and videos [12]. LLMs are a type of GenAI specialized in processing and generating text (including computer code). Most recent AI grading studies either solely rely on LLMs, or use a multi-modal model such as GPT-4 or GPT-4V that could process both text and images [13]. Since the current study only involves processing and generating text, in the rest of the paper we will use the term LLM instead of the term GenAI.

### A. Prior studies on LLM grading

A number of recent studies have tested foundational LLMs' grading capabilities across a range of different STEM topics, response types, grading outcomes and grading strategies. For example, Kortemeyer [7] used GPT-4 to grade short answers to science questions by two-way (True/False) or three way (Correct/Partly correct/Incorrect) outcomes, and studied whether providing a reference answer in the grading prompt is beneficial. Lee et al. [14] used GPT-3.5 and GPT-4 to grade students' short answers to science questions, and compared the effectiveness of six different prompting strategies using prompt templates with replaceable components. Henkel et al. [15] used GPT-3.5 and GPT-4 to grade short answers to K-12 science questions by correct or incorrect, and compared to the consensus of two human raters. Carpenter et al. used GPT-4 [16] to grade short, single step answers to 13 university level

*Contact author: Zhongzhou.Chen@ucf.edu

computer science questions, categorizing the answers into Correct, partially correct and incorrect.

For grading longer student answers to more complicated multi-step problems, Sonkar et al. [5] compared the performance of multiple language models on grading longer responses for university level Chemistry questions. Golchin et al. [17] used GPT-3.5 and GPT-4 to grade responses in several Massive Open Online Courses, with topics ranging from History to Astronomy, and compared LLM grading to both instructor and student peer grading. Liu et al. [13] used GPT-4V to grade students' written mathematical derivation in a university level math test according to a multi-item rubric.

For university level physics, Yan, Zhang and Jia graded four university level conceptual physics questions using GPT-4, using rubric items scored on a scale of 0-3, and examined LLM grading accuracy with or without reference answers [18]. Kortmeyer and colleagues [19,20] usedGPT-4 or GPT-4V to grade students' hand-written responses in upper level or advanced intro level thermodynamics problems, and investigated a number of issues such as comparing different grading workflows, generating confidence indicators for grading outcome, to iteratively improving rubric item writing. Our own earlier study [21] compared the effectiveness of three different prompt styles on grading of one physics conceptual question.

### B. Common strategies to improve LLM performance

Most of those studies claim that LLMs, especially GPT-4, can reach human-level grading accuracy. To achieve this level of grading accuracy, three strategies are commonly used to improve LLM performance.

***Chain-of-thought (COT) prompting***: COT prompting [22] instructs the LLM to first generate the reasoning steps, then generate the final answer. A simple form of COT prompting could be achieved by simply appending "Let's think step-by-step" to the prompt [14]. COT prompting has been shown to significantly improve the performance of LLMs in reasoning tasks, as the reasoning steps generated improve the chances that the LLM would predict the correct end result. For LLM assisted grading, it is also common to explicitly instruct the LLM to first compare students' response to the grading rubric, using language such as "`Please compare the student response with grading rubric`" [23] or "`The student says X. The rubric states Y. Based on the rubric, the student earned a score of Z.`" [24]

***Few-shot and zero-shot learning:*** Few-shot learning means that the LLM learns how to perform a task from a small number of examples, usually provided as part of the prompt [25]. In the context of LLM assisted problem grading, few-shot learning means that a relatively small number (between about 1 – 20) of student response and corresponding gradings were provided to the LLM (for example, see [14,15,26]). In some studies, the examples were carefully selected to represent typical student answers [24]. Some studies have found that few-shot learning yields better performance compared to providing no examples (zero-shot learning) [14,16]. In contrast, zero-shot prompting tasks the LLM to perform a given task without any examples. Studies involving either longer answers, or problems across different domains are more likely to adopt this strategy, which may be due to the difficulty of creating large numbers of quality examples [5,13,17].

***Self-consistency:*** A less often used technique is self-consistency [27], which is sometimes also referred to as ensamble-voting [14]. The self-consistency strategy simply tasks the LLM to perform the same task multiple times, (usually between 3-10 times), and take the most common result (most self-consistent) as the outcome. The self-consistency strategy increases performance by utilizing the inherent randomness of LLMs, and mitigates the negative impact of occasional hallucination or erroneous reasoning by a voting mechanism. Lee et al. compared ensamble-voting strategy outcome to individual grading outcomes, but found that for GPT-3.5 and GPT-4 models, the technique did not significantly improve the grading accuracy [14]. Kortemyer et al. employed a similar strategy for continuous point outcome, by taking the mean grade from 10 runs [19].

### C. Underexplored Areas in LLM grading

Although the above mentioned studies have explored multiple different grading strategies, there are still several areas of LLM grading that have not been fully explored.

First, content-wise most studies used LLMs to grade either short, verbal response to one or two step simple conceptual problems, or math derivations for multi-step numerical calculation problems [7,13,19,28]. Few studies have asked students to verbally explain their reasoning process for multi-step calculation problems, and use LLMs for partial credit grading. For introductory level physics, the instructional focus is more on teaching students to "know which equation to apply", rather than manipulating the equations to reach the final answer. Therefore, assessing students' reasoning during problem solving is highly valuable. Second, there were not many studies that systematically explored the potential of prompt-engineering [29] for improving the performance on LLM graders. Most existing studies used one or two prompts, and compared strategies such as whether a grading rubric should be given [5,17], rather than comparing different methods of writing a prompt.

*Contact author: Zhongzhou.Chen@ucf.edu

Most prompts used for LLM grading contain both grading instructions and grading rubric, both of which could be optimized to improve performance. For grading instructions, while COT prompting technique is a popular choice, the actual format of COT prompting in different studies is quite different. Our own earlier study with a small set of data suggests that details of COT prompt writing may have significant impact on the grading outcomes [21]. Regarding grading rubrics, one recent study [20] examined iteratively refined rubric item writing using Item Response Theory, but did not find significant improvements in outcome. Our own preliminary study showed that adding detailed explanation to rubric items could improve grading performance for GPT-3.5 model [21].

In addition, self-consistency as a potential strategy for improving LLM grading outcome has not been employed and tested in many studies.

Beyond improving the accuracy of LLM grading, establishing an effective and equitable LLM grading procedure also requires careful consideration of how machine interface and integrate with humans, including enhancing human oversight of grading outcome and improving grading outcome transparency for students.

For enhancing human oversight, one study explored generating a confidence metric that could identify potentially incorrect grading cases. Liu et al. [13] used the standard deviation of continuous grades from multiple grading runs as a confidence indicator, and found that indicator has an average accuracy of 0.69 in identifying wrong grading. Whether a similar strategy could be applied to categorical or binary outcome grading cases have not been studied.

For increasing grading transparency for students, we argue that LLM grading assistances could be easily modified to generate effective student facing feedback. As pointed out in Lee et al. [14], output of COT grading prompt contains the rationale of LLM's grading decision, which makes the grading more understandable for human experts. This presents an opportunity to generate student facing feedback by modifying the COT grading output using LLMs.

### D. Research Questions and Design of Current Study

In order to further explore the areas mentioned above, the current study will seek to answer the following three research questions.

**RQ1:** Can LLMs approach human-level accuracy on grading students' reasoning of multi-step calculation problems, using a combination of prompt-engineering and self-consistency techniques?

**RQ2:** Can a grading confidence index be developed for binary outcome rubric, using the level of variation between multiple grading runs? How effective is the confidence index?

**RQ3:** Can LLMs be tasked to generate adequate quality student facing feedback based on the output of COT grading prompt?

To answer those questions, the current study collected data from three problems that were administered on two exams in two sections of a university introductory level physics course on Newtonian Mechanics. For each problem, students were asked to explain their problem-solving process in words (including math expressions), in addition to choosing or entering the correct final answer/number. The three problems consist of one conceptual problem and one numerical problem on the topic of conservation of mechanical energy, and a second numerical problem on conservation of linear momentum. Students' written responses were initially graded based on a multi-item rubric by one human-grader, and awarded partial credits for partially correct explanations. A second grader then graded all responses independently for research purposes.

We then conducted the following LLM grading experiments using both GPT-3.5 and GPT4o models:
1. We compared the grading performance of different versions of grading prompts, using prompt templates with replaceable prompt components similar to what was used in Lee et al. [14].
2. We compared the grading accuracies between a single grading run and the mode score from five independent runs using the self-consistency strategy.
3. We tested using the levels of variation between different self-consistency grading runs as a confidence metric for identifying potentially problematic grading cases.
4. We tasked LLM to generate student facing grading feedback, based on the outcome of COT grading, and asked two experienced instructors to evaluate the quality of the feedback.

We made the following four deliberate choices in designing our grading approach, which sets apart the current study from existing studies.
1. *Each rubric item in a multi-item rubric is graded as binary outcome (1/0).*

While binary outcomes are mostly used for single-step short answers in previous studies, we chose to grade each rubric item on a binary scale for multi-step problems because of three reasons. First, using a multi-item rubric on a binary scale instead of a single-item rubric on a multi-point scale can, in principle, lead to a more accurate test of the level of agreement between two grading outcomes as it examines each item separately. For example, for a three-item binary rubric, {1,0,1} and {1,1,0} are different by 2 digits, yet both has the same total score of 2. The two grading outcomes would have been considered as in agreement if graded on a single-rubric on a scale of 0-3. It is worth noting that Liu et al. [13] found that fine-grained binary rubric results in lower human-machine agreement compared to continuous holistic rubric, which could indicate that agreement on binary output rubric presents a greater challenge for LLMs than continuous scale outcome. Second, binary outcome makes it easier to find a mode grading outcome from a relatively small number of self-consistency runs. Compared to categorical outcome such as 0, 0.5, 1 (incorrect, partially correct, correct) for each rubric item, binary outcome for each item reduces the possible number of different grading outcomes, allowing a consensus to be found with relatively fewer numbers of runs.

*Contact author: Zhongzhou.Chen@ucf.edu

Finally, we believe that any continuous or categorical grading can in principle be reduced to multiple-item binary outcome, if the scores are consistent and fair across the entire student population. For example, consider a rubric item that is graded on an integer scale of 0-2. If the grading is fair for all students, then we can always compare the answers from all students who scored 1 to those who scored 2 (or to those who scored 1), and identify at least one common difference between those answers to justify grading. By doing so we have essentially created one new binary outcome rubric items that differentiates score 1 from score 2. Following this process one can reduce any continuous outcome rubric item into multiple binary rubric item. Therefore, multi-item binary rubric grading can be seen a more fundamental grading challenge.

2. *Grading rubric focuses on the reasoning behind problem-solving process, instead of the calculational fluency.*

First, since introductory level physics courses focus more on conceptual understanding than calculational fluency, grading rubrics for introductory level physics problems often assign the majority of partial credits to selecting and properly applying the physics principles to a problem, and less on correctly calculating the results. Therefore, grading the reasoning process alone could capture the majority of partial credits in most introductory physics problems. Second, since many introductory level students have limited experience with LaTeX, their math input is often non-standard and challenging to understand even for human graders. Furthermore, students have the freedom to define their own conventions in problem solving, such as deciding the positive direction of the y-axis to be either up or down. Third, evaluating math expression is more challenging due to at least two reasons. First, LLMs as stochastic systems are not always reliable for rule-based tasks such as mathematical manipulation or logical inferences [30]. Rather, accuracy of math expression can be much more efficiently evaluated using tools such as python or Mathematica.

3. *Ground truth of grading is established using the grading of two independent human graders, who are not tasked with reaching a grading consensus.*

Most previous studies took as ground truth the grading of either a single human grader [13,18], or the consensus reached by two/three graders [15,24]. In this study, we chose to use two sets of grading from two human graders for several reasons. First, using a single human grader's rating as ground truth can be problematic as human graders themselves are far from perfect [31]. Second, the level to which two human graders agree with each other serves as a key baseline for "human-level" grading consistency. In other words, the LLM grader can be seen as "good enough" when the level of agreement between LLM and human is comparable to the level of agreement between two human graders. Using a consensus of multiple graders does not present the baseline measure of grading consistency among humans. Finally, there are cases where the student responses are inherently ambiguous, and can be interpreted more than one viable way. As a result, in this study we use cases in which LLM grading differs from both human graders as an indicator for potentially incorrect grading, which is useful for evaluating the effectiveness of the confidence index and answering RQ2.

4. *No standard reference answer or example grading is provided.*

At least two previous grading studies have shown that providing standard reference answers could harm LLM's grading performance, especially for more complicated multi-step problems [7,18]. It was suggested that reference answers could limit LLM's ability to accept different forms of expressing the same correct idea. For a similar reason, we chose a zero-shot approach over a few-shot approach. In addition, providing grading examples has two significant practical limitations. First, it can be time consuming for human graders to select typical responses and write grading justification, especially for longer answers with COT prompting, in which the example must also contain step-by-step thinking process. Second, the examples will significantly increase the length of the prompt, and as a result significantly increase the cost of the total grading process. Finally, a zero-shot approach would make it more straightforward to isolate the performance improvement from prompt-engineering which enables us to answer RQ1.

*Contact author: Zhongzhou.Chen@ucf.edu

The remainder of the paper is structured as follows. The detailed study setup, problems, problem administration and human grading process is described in Section II A. Details of LLM grading, including different prompt design and implementation self-consistency strategy, as well as feedback generation process are described in section Section II B. Section II C. explains the details of data analysis process and evaluation metric for answering the three research questions. We present the results of the three grading studies in Section IV, and discuss the implications of the results in Section V and VI.

## II. METHODS

### A. Study Design and Materials
#### 1. Problems and Grading Rubrics

The three problems used in the current study are labeled Question 1, 2 and 3 respectively.

**Question 1** was a conceptual question on the topic of conservation of mechanical energy, as shown in Figure 1.

Swimmers at a water park have a choice of two frictionless water slides. Both slides drop over the same height $h$: slide 1 is straight while slide 2 is curved, dropping quickly at first and then leveling out. How does the speed $v_1$ of a swimmer reaching the bottom of slide 1 compare with $v_2$, the speed of a swimmer reaching the end of slide 2?

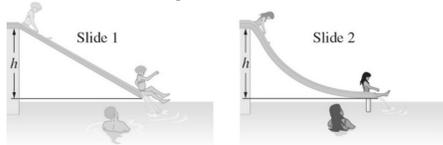

Figure 1: Problem text and figure of Question 1

Human graders used the following 3-item rubric when grading students' responses to Question 1:
```
Item 1: The student should mention either one
of the following:
  * conservation of energy OR
  * work and kinetic energy theorem
Item 2: The student mentioned either one of
the following:
  * No net external non-conservative work is
being done, so mechanical energy  is conserved
for the system  OR
  * the slide is frictionless/smooth OR
  * gravity is the only force that does work
on the girl.
Item 3: The student indicated either one of
the following:
  * potential energy is converted into
kinetic energy OR
  * Work done by gravity/gravitational force
is equal to the change in kinetic energy of
the girl
```

**Question 2** was a numerical input question which required students to input a numerical answer, on the topic of conservation of mechanical energy that involves two types of potential energy. This problem has two very versions, one with the ball launching from the spring tube (Figure 2), the other with the ball being dropped from a height h onto the spring tube. Students were presented with one of the two versions at random. Since the solution process of the two versions are nearly identical, we use a single rubric for grading the solutions of both versions.

A small massive ball of mass 0.46 kg is launched straight up from a tube using an ideal spring. The spring has spring constant k = 115 N/m and relaxed length of L0 = 0.74 meters. The spring was initially compressed to a length of L = 0.08 meters, and the ball was launched to a height of h = 0.76 meters above the top of the spring. Find the velocity of the ball at height h in units of meters per second. Retain your answer to two decimal places.

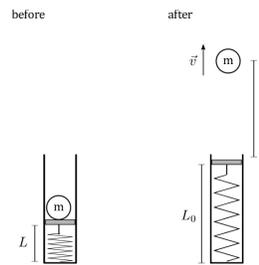

Figure 2: Problem text and figure of Question 2

Human graders assigned partial credit to student answers according to the following 3-item grading rubric:
```
Item 1: Wrote down conservation of mechanical
energy or indicated that mechanical energy can
be used to solve the problem
Item 2: The potential energy term of the
conservation of mechanical energy formula
contains both a gravitational potential energy
term and an elastic potential energy term.
Item 3: The gravitational potential energy
term contains an expression similar to (h + L
− L_0), and shouldn't be just mgh or mgL
```

Note that although it is possible to use work and kinetic energy theorem to solve this problem, no student chose that strategy due to the difficulty of calculating work done by the spring which requires integration. Therefore, the rubric was not designed to account for that.

**Question 3** was also a numerical input problem on the topic of conservation of linear momentum during a 2-dimensional collision (Figure 3). This problem was chosen because linear momentum is a vector concept, therefore the solution involves vector decomposition using trigonometry, which was not present in the previous two problems.

*Contact author: Zhongzhou.Chen@ucf.edu

Two icy boulders in Saturn's rings approach each other, collide, and stick together as shown in the figure below. The first has a mass of 145 kg and velocity of 18 $m/s$. The second has a mass of 360 kg and velocity of 10.7 $m/s$. The angle between the two velocities is $\theta = 26.5°$. Determine the magnitude of their velocity after they collide. Round your answer to the nearest 1 decimal place in units of $m/s$.

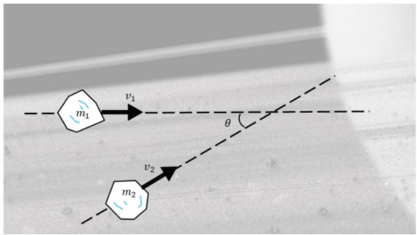

Figure 3: Problem and figure of Question 3

Human graders assigned partial credit to student answers according to the following 3-item grading rubrics:
`Item 1: Decomposed linear momentum of boulder 2 into its x and y components.`
`Item 2: Wrote down conservation of linear momentum equation for both the x and y directions independently.`
`Item 3: Used Pythagorean theorem to find the magnitude of the final velocity.`

For this problem, 10 students took an alternative correct solution path, which involves directly calculating the total linear momentum of the two boulders using the law of cosines, and dividing the result with the mass of both boulders. Since handling multiple solution paths is more complicated and out of the scope of the current study, we removed those student responses from the research data set.

### 2. Problem administration:

Question 1 was administered as part of a mid-term exam in one of the sections of an introductory mechanics course, and Questions 2 and 3 were administered as part of the final exam in another section of the same course. The enrollments for both sections were 99. Both exams were closed-book exams administered as a Quiz on Canvas learning management system [32]. Students took both the midterm exam and final exam in class with proctors present. Students were not made aware of the grading rubric during or before the exam.

For each question, students need to first either pick a choice item (Question 1) or enter a correct number (Questions 2 and 3), then provide a detailed explanation of the problem-solving process, including explaining the physics principle being applied. The explanation can be typed in a text input box on Canvas, which has a limited capacity for accepting mathematical symbols.

Students were instructed to explain their step-wise problem-solving process, including the physics principles involved, justification of the principles used, known and unknown variables, and other key steps. The detailed student facing instructions for explanation can be found in Appendix D.

### 3. Human grading of responses

For all three problems, one human grader initially graded only the responses from students who submitted an incorrect answer, and assigned partial credits to those students if their answer satisfied one or more rubric item. Each rubric was initially assigned a binary grade, and then multiplied by a factor determined by the instructor of the course based on the assigned point value of the problem. The amount of total partial credit available added up to about 70%-80% of the assigned point value of the problem.

For research purpose, the first human graders graded the remaining responses for students who submitted the correct answer. A second human grader also graded all the responses independent from the first human grader. For Question 3, a third human grader also graded all the responses, due to exceptionally low levels of agreement between the first two graders. The process is explained in more detail in the Results section.

### B. LLM grading and Feedback Generation

#### 1. Large Language Models

LLM grading was conducted using two chat models deployed on the Azure OpenAI platform: gpt-35-turbo version 0301, and gpt-4o version 2024-05-13. The model is instructed to operate at temperature of 0.8, which is similar to several other LLM grading studies [5,13,14]. The maximum output token is set to 1000. Gpt-4o was also used for feedback generation. We chose gpt-4o over gpt-4 since gpt-4o offers similar performance at 1/10 of the cost.

#### 2. LLM grading of student responses

Communication with LLMs was through the Azure Python API in the Langchain_openai package version 0.1.8 [33]. Grading of each response is a separate API call to the LLM, ensuring that the LLM do not have memory of previous gradings or other student responses. The grading output is stored and organized into python dataframes using the pandas package version 2.2.1 [34].

**2.1 Prompt Template**

The prompt provided to the LLM contains two messages. A system message that tells the model what roles to play and how to behave in general [35], and a human message that contains the student response and grading instructions. For this study the same system message is used for all grading tasks, while the human message changes on every grading case.

The system message used in this study is:
`You are a college introductory physics teacher who is grading a student's written explanation to a physics problem based on a grading`

*Contact author: Zhongzhou.Chen@ucf.edu

rubric. Your grading always ends with a comma
separated binary vector.

The human message (or the prompt) is constructed according to the following prompt template:
```
Here is a college introductory level physics
problem:
"{ProblemBody}"
Students are instructed to provide an
explanation to their answer.
Student explanations are being graded based on
the following rubric:
"{Rubric}"
Grading is performed strictly according to the
following requirements:
# The grading must start with the evaluation
of each individual rubric item.
{Requirements}
# For each rubric item, the student
explanation will receive 1 point if the
explanation satisfies the rubric, or 0 point
if the explanation does not satisfy the
rubric. Never assign a 0.5 for an item.
# Each rubric item is graded only once.
# Steps or sentences in student's explanation
may not follow the same order as the rubric.
# Conclude the grading response with vector of
length 3, included in curly brackets and
separated by commas, such as {{0,0,0}} or
{{1,0,1}} or {{1,1,1}}. The vector summarizes
the grading of each of the three rubric items.
Student response:
"{StudentResponse}"
Grading:
```

The template contains four replaceable components, enclosed in curly brackets, that can be changed programmatically between different grading tasks. The `ProblemBody` component contains the body of the problem, and is changed for each problem being graded. The `Rubric` component contains the rubric used for machine grading. The "`Requirements`" component contains different instructions on how the language model should respond to the grading task. Those two components are the target of prompt-engineering efforts. Finally, the `StudentResponse` component contains the actual student response to the problem to be graded.

**2.2 Rubric Design:**

We tested two different styles of rubric writing. "Simple rubric" is essentially the same as rubric given to the human graders, with added texts "Students could" in front of each rubric item.

"Detailed rubric" expands the simple rubric by providing the LLM with additional explanations, instructing the LLM on types of expression that could be seen as satisfying the rubric item. For example, the Detailed rubric used for Question 2 Item 1 is as follows:
```
# Item 1: The student wrote down conservation
of mechanical energy equation or indicated
that mechanical energy can be used to solve
the problem
  * The student could write mathematical
expressions such as MEi = MEf, ME_i – ME_f =
0, or KE_i + PE_i = KE_f + PE_f, or other
similar forms
  * Students could use terms such as
"Energy", or "Mechanical Energy".
```

All other rubric items used in this study can be found in Appendix A.

For Question 2 and 3, based on the initial grading outcome, different versions of Detail rubric were tested.

**2.3 Grading Requirements:**

We tested the following three types of requirements for the grading process:

*Basic Chain of Thought (COT):* The COT requirement largely mimics the most common COT strategies being used in different studies, which read as follows:
```
For each rubric item, first write step by
step reasoning on why or why not the student
explanation satisfies or contradicts the item.
Then assign a binary grade of either 0 or 1,
with 1 indicating the student explanation
satisfied the rubric item, and 0 otherwise.
```

One deliberate design choice is to ask the LLM to first write the reasoning and then assign a grade, which is different from many other practices. This design is based on the fact that LLM predicts following tokens based on all previous tokens, including the tokens that were previously generated. Therefore, if the reasoning is presented before the grading, the grade generated will be influenced by the reasoning test.

*Explicit Compare:* The explicit compare prompt provides more explicit instruction on how the step-by-step reasoning should be constructed, by asking the LLM to directly compare student explanation to the item and item descriptions:
```
For each rubric item, first compare student
explanation with the rubric item and the item
description, then conclude if the explanation
satisfies or didn't satisfy the rubric item.
Finally, assign a binary grade of either 0 or
1, with 1 indicating the student explanation
satisfied the rubric item, and 0 otherwise.
```

*Forced Compare:* The forced compare prompt explicitly requires the LLM to follow a rigorous template of reasoning, and identify the most relevant parts of student reasoning based its gradings on. In our earlier study [21], this type of prompt resulted in superior performance compared to conventional COT prompt using GPT-35 and completion API.
```
# For each rubric item, write the grading
statement strictly following the order of the
statements below:
  ## First, state one of the following two:
```

*Contact author: Zhongzhou.Chen@ucf.edu

```
    "For item <<item number>>, the rubric
states that <<quote from the rubric item
description>>. The most relevant parts in the
student explanation are <<direct quote or
quotes from student explanation>>.
    "For item <<item number>>, the rubric
states that <<quote from the rubric item
description>>. No part in the students'
explanation is relevant to the rubric"
  ## then state one of the following:
    "the student explanation is similar to
this part of the rubric description <<most
similar part of the rubric>>",
    "the student explanation and the rubric
description are very different"
    "the student explanation and the rubric
description are irrelevant"
  ## Finally, conclude with a binary score:
    "so the grade is 1"
    "so the grade is 0"
```

### 2.4 Prompt Styles:

In this study we refer to the different combinations of rubric design and grading requirements as different prompt styles. Since in this study we emphasize finding the best performing prompt style, rather than an extensive exploration of understanding what makes a prompt effective, we did not completely test all possible combinations of rubric design and grading requirements, but instead selected the following four most representative combinations listed in Table 1:

Table 1: Correspondence between prompt styles, rubric styles, and grading requirements.

| Prompt Style | Rubric | Requirement |
| --- | --- | --- |
| Simple COT | Simple | Basic COT |
| Detailed COT | Detailed | Basic COT |
| Detailed Compare | Detailed | Explicit Compare |
| Forced Compare | Detailed | Forced Compare |

### 2.5 LLM Grading and Error Detection:

The entire grading prompt including all its components was sent to the LLM for output generation. Upon receiving the output from the LLM, a python script examines whether the output contains a binary vector such as "{0, 1, 1}" (since all three questions happens to be graded on a 3-item rubric). If not, the exact same prompt was sent back to the LLM as a new grading task. The process was repeated until the output contained the intended vector outcome. In all the grading task in this study, the number of re-generated cases range from 3 – 10 per 100 responses graded.

### 2.6 Self-consistency grading

Each prompt style was first used once to grade all the student responses for each problem. Then the best performing prompt style, judged based on the evaluation metrics explained in the next section, was selected for self-consistency grading, which repeats the grading process five times. We chose 5 runs to allow sufficient detection of variation between different runs, while keeping the total run time for the gpt-4o model to under 2 hours. Self-consistency runs achieve three purposes:

1. To examine the stability of LLM grading between different runs.
2. To test if using the mode grading could achieve better performance than individual runs.
3. To test using the level of variation between different runs as an indicator for grading quality

The stability of individual runs is characterized by the average, minimum and maximum values of the key metrics for measuring the quality of grading outcome, explained in detail in the next section. Data tables containing those values can be found in Appendix B.

The level of variation between different self-consistency runs on the same response is measured using the Normalized Shannon Information Entropy (or entropy for short) between different runs, defined as [36]:

$$H_{REL} = -\frac{\sum_{i=1}^{n} p_i \log_2(p_i)}{\log_2(n)}$$

In the context of self-consistency grading of one student response, $p_i$ is the frequency with which each grading outcome as a binary vector (such as {1,0,1}) appears, and $n$ is the total number of runs ($n = 5$ in the current study). When each run gives a different grading outcome, $H_{REL} = 1$, when all runs give unanimous grading outcome, $H_{REL} = 0$.

For a particular response, mode score is defined as the most common grading outcome among all 5 runs, if there is more than one grading outcome. In the rare case of a draw, the first outcome to appear in the data table was selected. As a result, the entire set of mode scores for all the responses is in most cases different from the grading outcomes of any of the 5 individual runs.

### 2.7 Student Facing Feedback Generation and Evaluation

Student facing grading feedback is generated by sending the following information to the LLM: the problem body, the students' response, the rubric, the grading output of the top performing LLM grading prompt for each problem, and some instruction for feedback format. The grading output includes both the LLM generated reasoning text and the final grade. The format instruction instructs LLMs to write feedback that contain three parts:1. A summary of the grading rubric. 2. Feedback to students regarding how their solution satisfies the rubric items. 3. Ask students to contact their teacher if they have any questions.

Note that for all three problems, the mode-score for self-consistency grading always outperforms individual grading runs. However, for the current study, student facing feedback is generated based only on the top performing individual grading run to reduce complexity of the process. The full prompt for feedback generation, including system message, human message and changeable components, can be found in Appendix C. Initial testing on a few student responses showed that gpt-4o writes significantly better feedback than gpt-35, so for the current study we use gpt-4o to generate student facing feedback.

*Contact author: Zhongzhou.Chen@ucf.edu

A subset of the feedback messages was evaluated by two instructors. Both instructors hold PhD in physics education research and have taught introductory level Newtonian mechanics multiple times. We randomly selected 20 student responses and grading outputs for each question to generate feedback. The 60 generated feedback messages, together with the problem text and student responses, were given to the instructors to be rated on a scale of 1-5, according to the following explanation:

    1: I need to completely re-write the feedback before giving it to the student.
    2: I need to re-write most of the feedback before giving it to the student.
    3: I only need to make some minor changes before giving it to the student.
    4: I am fine with giving it to the student without making any changes.
    5: I really like the feedback and think that the student can benefit from it.

### C. Data Analysis:
#### 1. Metrics for grading quality.

In this study we graded student responses on a multi-item rubric, and graded each rubric item on a binary scale. Additionally, we chose the ground truth to be two separate human graders' performance. Therefore, the metrics used for measuring the agreement between two sets of grading outcomes are also somewhat different from other studies that grades on a continuous scale, and use either a single grader or consensus between two graders as ground truth. We use the following five metrics.

***Percent of matching grades***: The percentage of cases in which the LLM agrees with a human grader on *all* rubric items (i.e., the binary grade vectors must be identical).

***Average Simple Matching Distance (SMD):*** SMD between two binary vectors is defined as [37]:

$$\text{SMD} = 1 - \frac{\text{number of matching digits}}{\text{total number of digits}} = \frac{\text{number of mismatched digits}}{\text{total number of digits}}$$

SMD measures the difference between two binary grading outcome vectors. SMD = 0 when the two vectors are identical, and SMD = 1 when the two vectors are different on all items. SMD is a better choice for the binary grading task compared to the more popular Jaccard Distance, because SMD accounts for both matching grades of 0 and matching grades of 1, whereas Jaccard Distance only accounts for matching grades of 1. In binary grading, the information carried by a 0 grade is equivalent to that of the 1 grade.

In this paper, we measure the level of dissimilarity between two sets of grading using the average SMD, which is the SMD averaged over all cases (i.e., students) for each problem. Compared to the percent of matching grades, average SMD (or SMD for short) captures the contribution of partial matches.

***Percent of cases different from both human graders (Diff. Both):*** A third measure is the percentage of cases in which the machine grading outcome is different from the grading of both humans, including cases when the two human grading outcomes are the same or are also different. Cases where the LLM did not agree with either human on one or more rubric items indicate a more likely case of erroneous grading by the LLM rather than by the human. This metric turns out to be the most sensitive measure for small performance improvement in cases when the level of machine-human matching is already high.

***Macro F1 score:*** F1 score is a metric that is widely used in machine learning to assess the performance of a predictive model. For any binary classification, the F1 score is the harmonic mean of precision and recall [38]:

$$F_1 = \frac{2 \times \text{Precision} \times \text{Recall}}{\text{Precision} + \text{Recall}} = \frac{TP}{TP + \frac{1}{2}(FP + FN)}$$

Here, TP, FP, and FN stand for True Positive, False Positive, and False Negative, respectively. For the current multi-item rubric, F1 score is calculated for each rubric item for the entire data set, and Macro F1 score is simply the unweighted average F1 score for all rubric items. We use Macro F1 score instead of a weighted average F1 score since the weight of each rubric item can be treated as the same in all problems tested. One Macro F1 score can be calculated between machine grading and one human grader, and we use the average Macro F1 score of the two as an indication of machine grading performance.

***Quadratic Weighted Kappa (QWK)***: QWK is a metric widely reported in machine grading literature. QWK measures the agreement between raters beyond chance on ordinal data. The amount of partial credit a student earns on a problem can be treated as ordinal data by adding the binary score on all rubrics. However, since QWK only accounts for the total score, it is less accurate for binary grading of discrete rubric items, since it cannot account for the differences among cases such as {1,0,1}, {1,1,0} and {0,1,1}. We report the average QWK (QWK average between the two humans) just for making it easier to compare current results with existing literature on AI grading.

### 2. Data visualization:

In addition to the data tables, for each problem we visualize the three key metrics for grading quality in three line and dots plot: SMD, Diff. Both, and Macro F1. The SMD is plotted according to the average SMD between machine grading and two human raters.

On each plot, metric value of each individual grading run plus the self-consistency mode score are plotted as line connected dots, for gpt-35 and gpt-4o models separately. In addition, one or two separate dots are added to represent the average value of each metric from the five self-consistency runs. The error bars on those separate dots represent the maximum and minimum values of the metric among all five self-consistency runs.

### 3. Evaluating self-consistency entropy as confidence interval

We chose cases in which the LLM grading outcome vector differ from both human graders as potentially erroneous grading, and measured how often we could identify those cases by reviewing all grading cases with the self-consistency entropy higher than a threshold. In other words, we are treating entropy as a diagnostic test for potentially problematic grading cases.

There are many metrics that can be used to assess the performance of a binary diagnostic test [39]. In the current study we focus on three of those that are most relevant to a test of grading quality, namely, sensitivity, precision, and F1 score. The cases that are potentially problematic are used as the ground truth for those three metrics.

1. Sensitivity or True Positive Rate: is defined as $\frac{TP}{TP+FN}$, which in this case is the fraction of potentially problematic cases that can be identified by reviewing all the high entropy cases.
2. Precision is defined as $\frac{TP}{TP+FP}$, which can be seen as the fraction of cases that are potentially problematic among all high entropy cases.
3. F1 score: The F1 score is defined in the previous section. However, the current F1 score signifies the effectiveness of the diagnostic test, and shouldn't be confused with the Macro F1 score of grading accuracy.

### 4. Technic Detail of Data Analysis

All data analysis was performed using R version 4.3.3, the tidyverse package version 2.0.0 [40], and the rstatix package version 0.7.2 [41]. QWK scores were calculated using Metrics package version 0.4.1 [42], and F1 scores were calculated using MLmetrics package version 1.1.3 [43].

## III. RESULTS
### A. Human Grader Baseline

In Table 2, we list the number of responses for each question, as well as the four measurement metrics between the two human graders on each question as a baseline for comparison. Question 1 had the highest level of agreement between the human graders, while Question 3 had the lowest level of agreement. It is worth noting that Question 3 was initially graded by a grading assistant and an experienced instructor, with only 62.3% level of agreement and average SMD of 0.17. A second experienced instructor graded the solutions again, and reached 70.1% agreement with the first experienced instructor. Therefore, the grading of the grading assistant was not used in the following analysis. This suggests that accurate grading is more challenging for Question 3, which may be due to either the lower quality of student responses or less well-defined rubric items.

Table 2: Level of agreement between two human graders by measurement metrics.

| Problem | N | SMD | match | F1 | QWK |
|---|---|---|---|---|---|
| Question 1 | 96 | 0.087 | 74.0% | 0.925 | 0.908 |
| Question 2 | 83 | 0.108 | 72.3% | 0.816 | 0.825 |
| Question 3 | 77 | 0.130 | 70.1% | 0.858 | 0.869 |

The percentage of student answers satisfying each rubric item judged by each grader (graded as 1) is listed in Table 3. It should be noted that "h1" and "h2" are different human experts for each of the three problems. As can be seen from the table, for Question 1, item 1 is easier to satisfy than items 2 and 3. For Question 2, item 1 and 2 are quite easy to satisfy while item 3 was much more difficult. For Question 3, all three items were about the same. There were no substantial differences in percent correct between the two human graters for each problem.

Table 3: Percent correct of each human grader on each rubric item by each problem.

| problem | grader | item 1 | item 2 | item 3 |
|---|---|---|---|---|
| Question 1 | h1 | 56.2% | 28.1% | 38.5% |
| Question 1 | h2 | 55.2% | 30.2% | 36.5% |
| Question 2 | h1 | 81.9% | 77.1% | 26.5% |
| Question 2 | h2 | 84.3% | 73.5% | 24.1% |
| Question 3 | h1 | 49.4% | 50.6% | 50.6% |
| Question 3 | h2 | 59.7% | 54.5% | 59.7% |

### B. Cost and time of Grading tasks

*Contact author: Zhongzhou.Chen@ucf.edu

For each of the three problems, the cost of grading all student responses once using LLM ranged between $0.2 to $0.3 using GPT-3.5 model, and between $0.7 to $1.00 using GPT-4o model. The difference in cost between different prompt styles is negligible. The time for grading all responses for each problem is under 10 minutes for GPT-3.5 model, and between 20 to 25 minutes for GPT-4o model. The differences between different prompt styles were also not significant, with Simple COT being the fastest.

### C. LLM Grading Performance

**Question 1:** As shown in Table 4 and Figure 4, GPT-4o model significantly outperformed GPT-3.5 model on every measure under all prompt versions. Within GPT-4o model, both Detailed Compare and Detailed COT outperformed Simple COT. Detailed Compare reduced cases that differed from both human graders by 1 case (1%) compared to Detailed COT, and slightly improved the average Macro F1 score. Forced Compare resulted in significantly worse performance in all metrics. Based on those results, we chose Detailed Compare prompt to conduct the additional 5 self-consistency runs.

Self-consistency runs show that the slight advantage of Detailed Compare over Detailed COT prompting is within the range of possible fluctuations between different runs, so the difference between the two prompt styles is minimum. The mode score of all five self-consistency runs outperformed the original Detailed Compare grading run, and is slightly better than the best performing run of the five individual self-consistency runs (see Appendix B for data). Based on all five metrics, the level of agreement between self-consistency mode score and human graders were almost identical to the level of agreement between human graders.

It is worth noting that the grading using Detailed Compare prompt seems to agree with human grader 1 more than human grader 2, whereas detailed COT seems to evenly agree with both graders.

In summary, for question 1 we found that both Detail Compare and Detailed COT outperformed other two prompt strategies, and that Detailed Compare outcome was stable across multiple runs. Self-consistency mode grading outperforms individual Detailed Compare grading and reaches human level agreement with human graders.

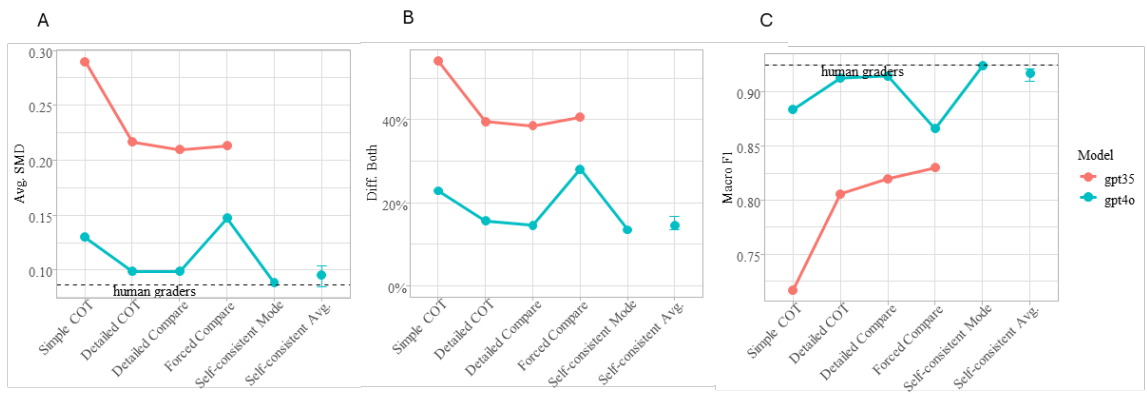

Figure 4: Outcome of different measures of agreement/disagreement for Question 1: A) Average SMD B) Different from both human graders. C) Average Macro F1 score. The unconnected dots on each graph represents the average value from five self-consistency runs, and the error bars correspond to the best and worst single runs. Level of agreement between two human graders are indicated using dotted lines for Average SMD and average Macro F1 score.

Table 4: Grading accuracy measures for LLM against a human grader for all prompt styles tested for Question 1. h1 and h2 indicate the two human graders. Macro F1 and QWK represent the average Macro F1 score and average QWK score from the comparisons with the two human graders, respectively. Asterisk indicates the best performing prompt on a single grading run.

| Model | Prompt_Style | SMD.h1 | match.h1 | SMD.h2 | match.h2 | diff.both | Macro F1 | QWK |
|---|---|---|---|---|---|---|---|---|
| gpt35 | Simple COT | 0.278 | 42.7% | 0.302 | 37.5% | 54.2% | 0.717 | 0.598 |
| gpt35 | Detailed COT | 0.212 | 56.2% | 0.222 | 54.2% | 39.6% | 0.806 | 0.730 |
| gpt35 | Detailed Compare | 0.201 | 55.2% | 0.219 | 52.1% | 38.5% | 0.820 | 0.736 |
| gpt35 | Forced Compare | 0.219 | 50.0% | 0.208 | 54.2% | 40.6% | 0.830 | 0.729 |
| gpt4o | Simple COT | 0.132 | 69.8% | 0.128 | 65.6% | 22.9% | 0.884 | 0.861 |
| gpt4o | Detailed COT | 0.097 | 74.0% | 0.101 | 74.0% | 15.6% | 0.913 | 0.917 |

*Contact author: Zhongzhou.Chen@ucf.edu

| Model | Prompt_Style | SMD.h1 | match.h1 | SMD.h2 | match.h2 | diff.both | Macro F1 | QWK |
|---|---|---|---|---|---|---|---|---|
| gpt4o | Detailed Compare* | 0.090 | 77.1% | 0.108 | 72.9% | 14.6% | 0.915 | 0.922 |
| gpt4o | Forced Compare | 0.153 | 65.6% | 0.142 | 62.5% | 28.1% | 0.866 | 0.848 |
| **gpt4o** | **Self-consistent Mode** | **0.080** | **78.1%** | **0.097** | **71.9%** | **13.5%** | **0.924** | **0.905** |

**Question 2:** As shown in Figure 5 and Table 5, the difference in performance between GPT-4o and GPT-35 is even greater than in Question 1. Within GPT-4o model, differences between different prompt styles were smaller, with Detailed Compare outperforming other two prompt styles on most metrics. Simple COT prompting had a very high level of agreement with grader h2, but a much poorer level of agreement with grader h1. The Macro F1 score was also slightly higher than the one with other prompt styles, likely due to high levels of agreement with grader h1. Performances by Detailed COT and Detailed Compare prompts were much more balanced. Forced compare prompts performed similarly to Detailed COT, but not as good as Detailed Compare. Therefore, we chose Detailed Compare prompt to conduct self-consistency runs.

Self-consistency runs show that the performance differences between different prompt styles were similar to the range of possible fluctuations between different runs. The mode score once again outperforms all the individual runs, and is closest to human level agreement, and even had identical Macro F1 score as between human graders.

However, self-consistency mode score still had different grading with both human graders on 19.3% of the cases, which is much higher than the best performing prompt for Question 1. When examining some of the cases where the LLM grader differs from both humans, we identified two main patterns. First, since the final elastic potential energy of the system was zero, some students omitted the term in their solution, stating for example "`I solved for MEi which is equal to kinetic energy plus potential energy … next I solved for the final mechanical energy which is the kinetic energy of the ball (1/2mv^2) and the potential energy of the ball at the height (m-gh)`". Yet, GPT was meticulously checking for the presence of both kinetic and potential energy in both initial and final states, giving grading justifications such as "`* However, the student does not mention elastic potential energy or provide expressions like 1/2k(L-L0)^2.`". Second, the original rubric explanation may have over emphasized that the gravitational potential energy should be "mg(h+L0-L)", and the AI grader seems to be checking for the presence of both L0 and L in the math expression. However, some students could have verbally described the solution process as "`However, for h I added the change in L as the ball traveled that distance in addition to the height from which h is representing.`". Others chose a different reference height for zero potential energy, resulting in "`mgh`" and "`mg(L0-L)`" on different sides of the conservation equation, which confuses the AI grader.

Therefore, we modified the rubric explanation by adding text explaining those two cases to the AI, which we refer to as "Detailed compare 2" (see Appendix A). As shown in Figure 5 and Table 5, those modifications to the rubric explanation significantly improves the performance of the LLM grader on every metric (Detailed Compare 2), especially in dropping the fraction of cases in which the AI differs from both human graders from 21.7% to only 9.6%. The LLM graders agrees with each human grader more than the human graders agree with each other, which means that in cases where the human graders disagree, the AI will most likely side with one of the two graders. The self-consistency runs with this new rubric (Self-consistent 2) shows that the improvements are stable. Once again, the mode score of the self-consistency runs had slightly better performance than all the individual runs.

*Contact author: Zhongzhou.Chen@ucf.edu

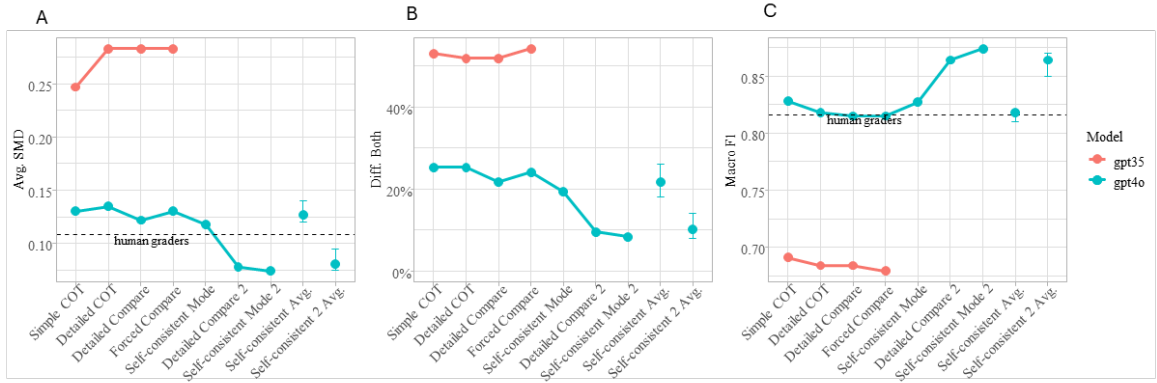

Figure 5: Outcome of different measures of agreement/disagreement for Question 2: A) Average SMD B) Different from both human graders. C) average Macro F1 score. The unconnected dots on each graph represents the average value from five self-consistency runs, and the error bars correspond to the best and worst single runs. Level of agreement between two human graders are indicated using dotted lines for Average SMD and average Macro F1 score.

Table 5: Grading accuracy measure for all prompt styles tested for Question 2. h1 and h2 indicate the two humangraders. Macro F1 and QWK are the average Macro F1 score and QWK score from the comparisons with the two humangraders.

| Model | Prompt_Style | SMD.h1 | match.h1 | SMD.h2 | match.h2 | diff.both | Macro F1 | QWK |
|---|---|---|---|---|---|---|---|---|
| gpt35 | Simple COT | 0.237 | 39.8% | 0.257 | 41.0% | 53.0% | 0.691 | 0.619 |
| gpt35 | Detailed COT | 0.281 | 38.6% | 0.285 | 39.8% | 51.8% | 0.684 | 0.487 |
| gpt35 | Detailed Compare | 0.281 | 38.6% | 0.285 | 39.8% | 51.8% | 0.684 | 0.487 |
| gpt35 | Forced Compare | 0.277 | 41.0% | 0.289 | 34.9% | 54.2% | 0.679 | 0.462 |
| gpt4o | Simple COT | 0.145 | 59.0% | 0.116 | 68.7% | 25.3% | 0.828 | 0.758 |
| gpt4o | Detailed COT | 0.129 | 63.9% | 0.141 | 63.9% | 25.3% | 0.818 | 0.734 |
| gpt4o | Detailed Compare* | 0.124 | 68.7% | 0.120 | 67.5% | 21.7% | 0.815 | 0.762 |
| gpt4o | Forced Compare | 0.124 | 66.3% | 0.137 | 63.9% | 24.1% | 0.815 | 0.726 |
| **gpt4o** | **Self-consistency Mode** | **0.112** | **69.9%** | **0.124** | **69.9%** | **19.3%** | **0.827** | **0.759** |
| gpt4o | Detailed Compare 2* | 0.080 | 80.7% | 0.076 | 79.5% | 9.6% | 0.864 | 0.878 |
| **gpt4o** | **Self-consistency Mode 2** | **0.064** | **81.9%** | **0.084** | **77.1%** | **8.4%** | **0.874** | **0.877** |

**Question 3:** As shown in Table 6 and Figure 6, GPT-4o still outperformed GPT-35 in all conditions tested and across all metrics. However, within GPT-4o, Simple COT had superior performance compared to other prompt styles, which is opposite from what was observed from the other two questions. The simple COT grading result even surpassed human level agreement on the average SMD and average Macro F1. However, when self-consistency runs were performed using simple COT prompt, we observed a much bigger fluctuation in outcome quality compared to the self-consistent runs for the two other questions. The outcome of different runs ranged from far superior than all other prompt styles, to comparable or even worse than other prompt styles. The original Simple COT happened to be one of the best of all six runs simply by chance.

However, the mode score (Self-consistent Mode) of the self-consistency runs with simple COT still outperformed the other three prompt styles. The average scores of the self-consistency runs on all metrics were also better than the other three prompt styles as shown in Figure 6.

To investigate why for this question, simple COT seem to sometimes outperform all other prompt styles, and whether it is possible to surpass the performance of simple COT, we tried three more grading prompts:

**Detailed Compare 2**: Adding more explanations to the existing detailed rubric, by studying cases where Detailed Compare (initial run) differed from both human graders.

**Simple Compare**: Using the simple rubric but with the "explicit compare" grading requirement.

**Detailed Compare 3**: Remove the original rubric explanations and replaced with explanations based on examining cases where simple COT grading (initial run) differed from both human graders.

*Contact author: Zhongzhou.Chen@ucf.edu

Detailed Compare 2 was designed based on similar strategy Detailed Compare 2 in Question 2. Both Simple Compare and Detailed Compare 3 prompt tests the hypothesis that the original instructor written rubric explanation actually reduced grading performance by over emphasizing certain forms of expression over other equivalent forms, causing the LLM grader to lose flexibility in handling a variety of student responses.

For Detailed Compare 2, one new explanation was added to rubric item 2, stating that students could "`write down one conservation of linear momentum equation, and either immediately or later indicate that this equation is applied to both x and y directions`". Two new explanations were added to rubric item 3, the first one stating that "`The Pythagorean theorem must by applied to the components of velocity or momentum.`", while the second one states that "`Simply stating "obtain the magnitude" of the velocity do not satisfy this rubric.`"

For Detailed Compare 3, rubric item 1 contains one explanation: "`The student could also write m2v2x and m2v2y in conservation of linear momentum equations without explicitly decomposing the momentum or the velocities.`" Item 2 contain two explanations, the first one stating that students can state linear momentum conservation on x and y directions separately, and don't necessarily need to write down the trigonometry, while the second one states that "`Only saying "momentum equation" or "used momentum" does not satisfy this rubric.`". Item 3 contain one explanation, stating that "`Just saying "put them together" does not satisfy this rubric`".

As can be seen from Table 6 and Figure 6, all three new prompts outperformed the original Detailed Compare prompt by a substantial margin on all metrics. When compared to the self-consistency mode score (using Simple COT prompt), all three prompts had similar average SMDs and lower fractions of cases different from both human graders. However, the Macro F1 scores are all lower than that of initial self-consistent mode, probably due to the grading outcomes are more similar to that of human grader h2 than human grader h1. Within the three prompts, the performance of Detailed Compare 3 had the lowest average SMD and lowest fractions of different from both humans, but also slightly lower Macro F1 score. Overall, the differences between the three prompt styles, especially between Simple Compare and Detailed Compare 3, are relatively small.

Based on those results, we conducted a second self-consistency run, this time using the Detailed Compare 3 prompt due to its small performance advantage over other prompts. For convention consistency, we named this run "Self-consistent 3", since it uses "Detailed Compare 3" prompt style. The mode score of self-consistent 3 outperformed almost all other grading methods tested by wide margin on all metrics, including self-consistency 1 (using simple COT). It even outperformed the original simple COT run, which was an exceptionally high performing run by chance. Self-consistent 2 mode score also clearly exceeded the human-to-human baseline in both average SMD and Macro F1 score. Moreover, the variance between the five runs in self-consistent 3 is much smaller than that of self-consistent 1, indicating that Detailed Compare 3 is a much more stable prompt than Simple COT.

In summary, on Question 3 we made the following observation: 1: The grading outcome of Simple COT prompt can have much wider fluctuations between different grading runs compared to Detailed Compare prompting. 2: Too much rubric explanation could hurt grading performance in certain cases, especially when simple rubric is already working well. 3. Rubric explanation that targets the errors in LLM grading under simple rubric significantly improves the grading outcome, whereas rubric explanation written by human expert alone have minimum improvement for performance.

*Contact author: Zhongzhou.Chen@ucf.edu

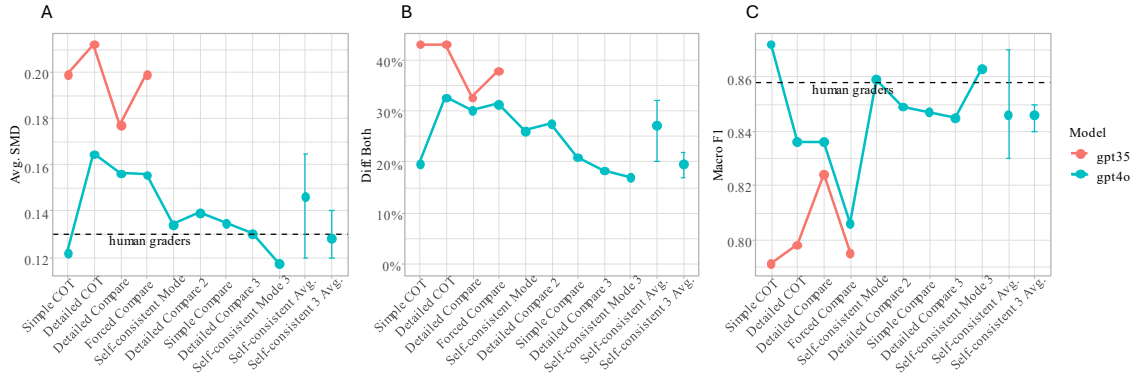

Figure 6: Outcome of different measures of grading agreement for Question 3: A) Average SMD B) Different from both human graders. C) Average Macro F1 score. The unconnected dots on each graph represents the average value from five self-consistency runs, and the error bars correspond to the best and worst single runs. Level of agreement between two human graders are indicated using dotted lines for average SMD and average Macro F1

Table 6: Grading accuracy measure for all prompt styles tested for Question 3. h1 and h2 indicate the two human graders. Macro F1 and QWK are the average Macro F1 score and QWK score from the comparisons with two human graders.

| Model | Prompt Style | SMD.h1 | match.h1 | SMD.h2 | match.h2 | diff.both | Macro F1 | QWK |
|---|---|---|---|---|---|---|---|---|
| gpt35 | Simple COT | 0.212 | 48.1% | 0.186 | 49.4% | 42.9% | 0.791 | 0.806 |
| gpt35 | Detailed COT | 0.221 | 48.1% | 0.203 | 49.4% | 42.9% | 0.798 | 0.750 |
| gpt35 | Detailed Compare | 0.177 | 57.1% | 0.177 | 57.1% | 32.5% | 0.824 | 0.790 |
| gpt35 | Forced Compare | 0.195 | 53.2% | 0.203 | 54.5% | 37.7% | 0.795 | 0.768 |
| gpt4o | Simple COT* | 0.143 | 67.5% | 0.100 | 75.3% | 19.5% | 0.872 | 0.877 |
| gpt4o | Detailed COT | 0.156 | 61.0% | 0.173 | 54.5% | 32.5% | 0.836 | 0.845 |
| gpt4o | Detailed Compare | 0.165 | 62.3% | 0.147 | 61.0% | 29.9% | 0.836 | 0.831 |
| gpt4o | Forced Compare | 0.177 | 54.5% | 0.134 | 64.9% | 31.2% | 0.806 | 0.847 |
| **gpt4o** | **Self-consistency Mode** | **0.160** | **63.6%** | **0.108** | **70.1%** | **26.0%** | **0.859** | **0.867** |
| gpt4o | Detailed Compare 2 | 0.152 | 61.0% | 0.126 | 66.2% | 27.3% | 0.849 | 0.854 |
| gpt4o | Simple Compare | 0.156 | 66.2% | 0.113 | 72.7% | 20.8% | 0.847 | 0.854 |
| gpt4o | Detailed Compare 3* | 0.165 | 61.0% | 0.095 | 77.9% | 18.2% | 0.845 | 0.871 |
| **gpt4o** | **Self-consistency Mode 3** | **0.143** | **66.2%** | **0.091** | **76.6%** | **16.9%** | **0.863** | **0.877** |

### D. Self-consistency entropy as an indicator of grading quality.

The effectiveness of using the self-consistency entropy as the grading quality indicator is listed in Tables 7, 8 and 9, for the three questions respectively.

For Question 1, 27.1% of all responses had entropy > 0, which included 69.2% of the cases where LLM grading differed from both human graders (sensitivity). If the entropy cutoff was raised to > 0.4, then only 17.7% of the grading responses requires review, which would identify 46.2% of all cases where LLM grading differed from both humans. Given that when using self-consistency mode score, only 13.5% of all cases differed from both human graders, using entropy > 0.4 as a confidence indicator would result in only about 7% of all students receiving potentially problematic partial credit grading without a human review.

Table 7: Effectiveness of using self-consistency entropy to identify problematic grading cases for Question 1.

| entropy | review | sensitivity | precision | F1 |
|---|---|---|---|---|
| >0 | 27.1% | 69.2% | 34.6% | 0.462 |
| >0.4 | 17.7% | 46.2% | 35.3% | 0.400 |

For question 2, with entropy > 0.4, under self-consistency 1 using the original detailed rubric, the fraction of response that require review (16.9%) is similar to that of Question 1, but the sensitivity is about 10% lower. Under self-consistency 2, which uses the improved detailed rubric, the fraction of cases that require review drops to 10.8% when considering cases with entropy > 0.4, while the sensitivity increased to 43%. Given that only 8.4% of the cases were potentially problematic, the method leaves only about 5% of potentially questionable grading without a human review.

*Contact author: Zhongzhou.Chen@ucf.edu

Table 8: Effectiveness of using self-consistency entropy to identify problematic grading cases for Question 2. SC1 and SC2 stand for self-consistency 1 and self-consistency 2, respectively.

| Run | entropy | review | sensitivity | precision | F1 |
|---|---|---|---|---|---|
| SC1 | >0 | 37.3% | 75.0% | 38.7% | 0.511 |
| SC1 | >0.4 | 16.9% | 31.2% | 35.7% | 0.333 |
| SC2 | >0 | 21.7% | 57.1% | 22.2% | 0.320 |
| SC2 | >0.4 | 10.8% | 42.9% | 33.3% | 0.375 |

For Question 3, under self-consistency 1 which uses simple COT as prompt, the fraction of response that requires review is much higher compared to the other two questions, which indicates that the grading outcome of simple COT prompt has greater variation between different runs. On the other hand, the sensitivity, precision and F1 score of the test are also higher than the other questions. For self-consistency 2 which uses detailed compare 3, the fraction of responses with high entropy is about 10% lower, similar to that of the other two questions. However, the sensitivity, precision and F1 scores are also lower. Since the LLM grading contains 16.9% of potentially problematic grading cases, a sensitivity of 38.5% would result in roughly 10% of students receiving potentially problematic grading.

Table 9: Effectiveness of using self-consistency entropy to identify problematic grading cases for Question 3. SC1 and SC3 stands for self-consistency 1 and self-consistency 3.

| Run | entropy | require review | sensitivity | precision | F1 |
|---|---|---|---|---|---|
| SC1 | >0 | 41.6% | 85.0% | 53.1% | 0.654 |
| SC1 | >0.4 | 23.4% | 60.0% | 66.7% | 0.632 |
| SC3 | >0 | 31.2% | 69.2% | 37.5% | 0.486 |
| SC3 | >0.4 | 16.9% | 38.5% | 38.5% | 0.385 |

Overall, the common pattern across all three questions is that when conducting 5 self-consistency runs on Detailed Compare prompting, entropy is a moderately effective indicator for potentially problematic AI grading. When the minimum entropy is set at 0.4, a human instructor will need to review about 10% - 17% of all responses, and end up detecting about 40% of potentially problematic grading cases. In all three cases, a higher entropy cutoff than 0.4 results in significantly worse sensitivity. Given that the top performing AI grader will generate only about 10% – 15% potentially problematic grading, we think that detecting 40% of those cases is at least acceptable in practice. Using entropy > 0 does improve the sensitivity of detection, but also requires the instructor to review a significant fraction of responses, which may not be worthwhile.

### E. Quality of Student Facing Feedback.

The instructor rating of AI generated student facing feedback is summarized in Figure 7.

Instructor h2 rated all feedback messages 3 or above (give to students with only minor modification), whereas instructor h1 rated 87% of the feedback messages 3 or above, and 8 feedback messages were rated with a 1 or 2. Of the 8 feedback messages rated with a 1 or 2, 5 was from Question 1 and 3 from Question 3. Six out of those 8 feedback messages were for response that received 0 or 1 out of 3 possible points. Upon closer examination, in most of those 8 cases the feedback did not specifically address students' reasoning. For example, one student response for Question 1 is:

```
"Assuming the kids travel the same distance
and they weigh the same, their speeds would be
equal when they reach the end of the slides.
Their initial and final positions are the
same. The displacement is the same. The height
is also the same. Therefore their speeds will
be equal.
K=(1/2)mv^2
p=mv"
```

This response was graded as 0 out of 3, and the AI generated feedback, after explaining the grading rubrics, only contains the generic statement of:

```
Unfortunately, I don't think your
explanation mentioned any of the aspects
listed above.
```

In contrast, below is an example from the same question, graded as 0 out of 3, and rated 4 by both instructors. This student wrote the following in the explanation:

```
1. Centripetal Force, due to the curve
of the second slide.
2. As stated previously, the second
slide is curved while the first is a
straight course into the pool.
3. When taking the curved state of one
pool and the straight course of the
other, the building of kinetic energy
in the second pool is greater as the
curve allows for a more sudden drop
and higher velocity to the end of the
pool.
```

This time the AI generated feedback that both directly addresses the students' response, and also briefly mentioned the correct solution method:

```
Unfortunately, I don't think your explanation
mentioned any of the aspects listed above. You
focused on centripetal force and the building
of kinetic energy due to the slide's
curvature, which are not relevant to solving
```

*Contact author: Zhongzhou.Chen@ucf.edu

```
this problem. The correct approach involves
understanding that mechanical energy is
conserved because the slide is frictionless,
and potential energy is converted to kinetic
energy irrespective of the slide's shape.
```

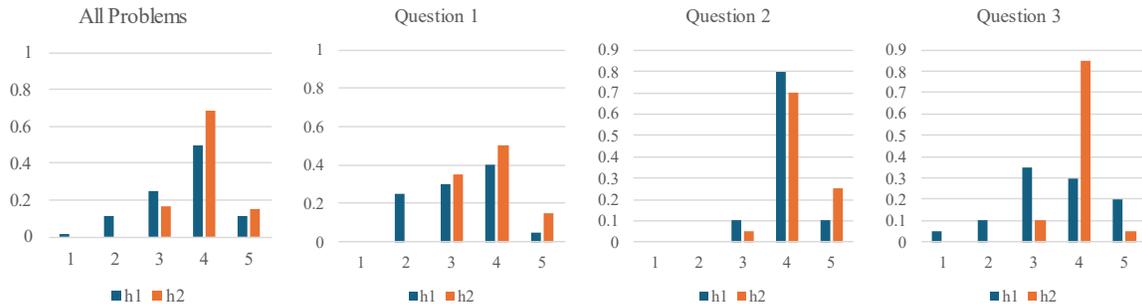

Figure 7: Distribution of instructor rating of the quality of LLM generated feedback messages. The two raters were labeled as h1 and h2.

## IV. DISCUSSION AND FUTURE DIRECTIONS
### A. Summary of Key observations

We first summarize the key observations from the study below, and compare them to results from existing studies.

1. For RQ1, providing the right amount and right type of rubric item explanation brings the most improvement to LLM performance, allowing GPT-4o model to agree with human raters in 70% - 80% of all cases, which is identical to or even higher than the level at which human graders agree with each other. Too few or too much explanation will both reduce grading accuracy. Explanation targeting errors in LLM grading leads to the most significant performance improvement.

*Explicit compare* grading instruction may have resulted in a small improvement in grading accuracy compared to *basic COT* instruction. Whether it also leads to smaller variation between different runs needs to be further tested in future studies. *Forced compare* instructions, on the other hand, significantly reduces performance for GPT-4o models, unlike what was observed for gpt-3 model in our own earlier study [21].

Adding rubric item explanation specifically designed for LLM grading has not been studied in previous LLM grading research. Compared to few-shot learning, adding rubric explanation is more cost-effective for improving grading accuracy, as it results in significantly shorter prompts. Furthermore, rubric explanation can be written to cover multiple possible situations in a few sentences, whereas each example in few-shot prompting only covers one situation each example.

2. The mode-score of 5 self-consistency runs outperforms the best single grading run in almost all cases. Combined with observation 1, those results show that a combination of prompt-engineering and self-consistency technique allows GPT-4o to achieve human level grading performance with zero-shot. Note that the mode-score performance is only significantly better than individual runs in Question 3, but not Questions 1 and 2. This might explain why Lee et al. [14] failed to observe significant performance improvement using a very similar strategy in their study, as their study involves simpler and shorter answers.

3. Utilizing rubric explanation and self-consistency, GPT-4o can grade written responses and generate good quality student facing grading feedback, at the cost of about $5 per 100 student responses. With Macro F1 score and QWK score both above 0.8, the grading performance of our best performing prompt is on par with most previous studies that claimed to have reached human-level grading accuracy (for example in [24]).

4. For RQ2, the entropy of self-consistency runs could serve as a moderately effective indictor, that allows a human instructor to identify 40% - 50% of potentially problematic grading cases, by reviewing 10%- 20% of all grading cases. The accuracy is lower than what was reported in Liu et al. [13], which might be partly due to the already high grading accuracy of self-consistency grading in the three problems in this study.

5. For RQ3, we verified that the student facing grading feedback can be straightforwardly generated by LLM based on COT grading outcomes. The vast majority of the feedback messages were highly rated by experienced instructors.

*Contact author: Zhongzhou.Chen@ucf.edu

### B. Advantages of LLM assisted grading over conventional human grading.

The above results demonstrated four significant advantages of LLM assisted grading over conventional human grading.

First, LLM-assisted grading is significantly cheaper and faster than human grading. This has been shown by many previous studies of AI-assisted grading. The current study further shows that even under the more stringent condition of binary outcome rubric items, GPT-4o can still perform at human-grader level accuracy for a fraction of the cost.

Second, LLM assisted grading has high levels of grading consistency, when employing item explanation and self-consistency, as seen in the small variations in self-consistency runs. Consistency of human graders on the other hand, are subjected to many factors such as fatigue and experience. We also observed in the case of Question 3 that specific human graders can sometimes have a low level of agreement with each other on the same rubric.

Third, LLM graders can continuously self-improve and accumulate grading experience, through either adding rubric item explanation, or via more sophisticated techniques such as Retrieval Augmented Generation (RAG) [44]. As shown in this study, refinement to rubric explanation based on previous grading outcome brings the most significant performance improvement. On the other hand, improving the accuracy of grading or understanding of rubric of human graders takes much longer time, and the experience usually does not carry over from one grader to the next.

Last but not least, LLM assisted grading can provide high levels of grading outcome transparency in the form of student facing feedback. It would be impossible for any human grader to write targeted feedback for each student response in any reasonable amount of time. Giving students transparent feedback is particularly important since even the best performing LLM grader with human oversight will still contain about 5 – 10% of potentially erroneous grading, and grading feedback provides students with an opportunity to raise different opinions about the received grade to the instructor. However, those advantages by no means imply that LLMs will or should completely replace humans in the grading process. On the contrary, our results suggest that input and oversight from one or more human experts is indispensable in the process of LLM assisted grading, and likely cannot be replaced even by future LLMs.

### C. The role of human-expert in LLM assisted grading.

Using the current LLM assisted grading process, human-expert input is indispensable for the following three steps:

1. Designing the initial grading rubrics. Items in the grading rubric indicate the aspects of the problem-solving process for which student in the current class should be evaluated against, and needs to be determined by a human-expert based on students' level of subject understanding. For example, whether students should be required to state the justification for using conservation of mechanical energy in a problem solution should be the decision of a human expert depending on students' level of proficiency of the concept. While part of the grading rubric authoring task might be automated by AI in the future, the decision-making process should always involve a human-expert who has experience with the student population.

2. Writing rubric item explanation for LLMs. Rubric explanation reflects the acceptable ways that an idea or a reasoning step can be expressed in the solution by a student. For example, writing "`1/2mv^2 = mgh`" might be seen as an indication that the student is implicitly considering the idea of conservation of mechanical energy, in the context of college introductory level physics. We demonstrated that this step is crucial in improving the agreement between LLM grading and human grading.

One possible explanation for why rubric item explanations are critical for LLM grading but not for human graders is that human graders tries to simulate the reasoning process of a student based on students' explanation during grading, whereas LLMs focus more on judging the semantic distance between the student response and rubric items. In the above example, the math expression "`1/2mv^2 = mgh`" could be semantically quite distant from the words "conservation of mechanical energy" on the surface level. However, a human grader would deduct that the student most likely had thought about energy conservation before writing down the expression.

While it may be possible to first instruct an LLM to generate simulated students' reasoning steps based on the written solution, then grade based on the generated reasoning steps, the entire process is likely significantly less efficient and less accurate compared to adding rubric item explanations. Furthermore, rubric item explanation also reflects the instructors' choice and requirement regarding acceptable types of expression. Therefore, we argue that writing rubric item explanation is a key step that necessitates input from a human expert.

3. Evaluating potentially problematic grading cases. A human expert is required to evaluate both the grading cases suggested by

*Contact author: Zhongzhou.Chen@ucf.edu

LLMs according to confidence indices such as entropy from self-consistency runs, and those raised by students after students have reviewed the grading feedback. The human expert could then improve grading prompt to reduce erroneous grading cases in the future.

The above three steps all involve judgements of experienced human instructors, and are unlikely to be replaced even my more capable models in the foreseeable future.

### D. Limitations and future directions.

Since the current study represents an initial exploratory effort in many areas of LLM assisted grading, it inevitably has some major limitations that would require a more systematic study in the future.

First, the current study only investigated problems with a single dominant solution. For problems that have more than one frequently appearing possible solution, a single grading prompt with multiple sets of rubric items may not be the optimal approach for grading. A more likely approach is to construct a multi-agent system, in which the first GenAI agent is tasked to identify which (if any) possible problem-solving method the student used, and sends both the student response and the closest problem solving method to the next GenAI agent to grade, according to the rubric designed for the particular problem-solving method.

Second, we deliberately chose not to evaluate the accuracy of students' mathematical fluency. How to accurately evaluate both problem solving logic and mathematical fluency in the same solution that contains both verbal response and math expression is a challenge that needs to be solved in future studies.

Nor did the current study explore grading problems or student answers involving graphs or diagrams, which require multi-modal models such as GPT-4V [13,20]. Future studies should also examine the reliability of the current method on different kinds of problems across different domain areas.

Third, the quality of student facing feedback could be further improved in future studies. The current feedback only focuses on grade-transparency, which explains why or why not students' answer satisfies a certain grading rubric. It is possible to provide specific feedback to students' common preconceptions, by providing LLMs with knowledge from physics education research (for example see [26]). That would make the grading feedback more beneficial for student learning.

Fourth, the quality and structure of student explanation likely can have significant impact on the performance of both LLM and human graders. In the current study, we provided detailed instructions to students on how to explain their reasoning. Future studies could investigate whether grading could be further improved by teaching students to write more clear and organized explanations of solution process.

On a related issue, future studies of AI-assisted grading should pay attention to whether and how students' response changes, as AI grading becomes more common and students are aware of the practice. In particular, whether students will be more likely to try to "trick" AI grader into giving partial credit by writing certain key words or template phrases.

Last but certainly not least, more research is needed to carefully examine the fairness of LLM-assisted grading, and study whether inherent biases of LLMs [31,45,46] impact grading outcomes for students with different backgrounds, especially different levels of English proficiency or language habits. For example, do responses produced by English as Second Language students receive lower grades from AI grading compared to native English speakers? How does the level of bias in LLM grading compare with that in traditional human grading?

### V. CONCLUSION

In this study, we experimented with using LLMs to grade students' verbal explanations of their reasoning process for complex, multi-step problems, according to a multi-item, binary outcome rubric. The results show that providing additional explanation to each rubric item could allow LLMs to consistently perform at or above typical human graders, especially when the explanations specifically target the errors of LLM grading. Furthermore, we showed that self-consistency is a useful strategy for both improving grading performance, and generating a confidence index for LLM grading. Finally, LLMs can easily generate student facing grading feedback to improve grading transparency. Overall, the study demonstrated that LLMs can be highly beneficial for grading large amounts of student written responses for introductory level physics problems.

### ACKNOWLEDGMENTS
This study was supported by NSF HER-1845436, and by UCF Pegasus Innovation Lab Digital Course Improvement program.

*Contact author: Zhongzhou.Chen@ucf.edu

*Contact author: Zhongzhou.Chen@ucf.edu

*Contact author: Zhongzhou.Chen@ucf.edu

# Appendix A: LLM Grading Rubrics

Question 1:
**Simple rubric:**
```
# Item 1: The student should mention either one of the following:
  * conservation of energy OR
  * work and kinetic energy theorem
# Item 2: The student mentioned either one of the following:
   * No net external non-conservative work is being done, so mechanical energy  is conserved for the system  OR
   * the slide is frictionless/smooth OR
   * gravity is the only force that does work on the girl.
# Item 3: The student indicated either one of the following:
   * potential energy is converted into kinetic energy OR
   * Work done by gravity/gravitational force is equal to the change in kinetic energy of the girl
```

**Detailed Rubric:**
```
# Item 1:
      "* The student should mention either one of the following in the explanation:
            ** Conservation of energy/mechanical energy. Conservation of Energy can be expressed in mathematical forms such as mgh = 1/2 mv^2, mghi+ ½mvi^2=mghf+ ½mvf^2, or MEi = MEf
            ** work and kinetic energy theorem.
            ** The student could explicitly mention both (gravitational) potential energy and kinetic energy, or mention both work and kinetic energy, without explicilty saying conservation of energy or name of the theorem.
      * Only mentioning the term potential energy will NOT satisfy this rubric.
      * The explanation cannot mention momentum, linear momentum, or centripetal forces"

# Item 2:
      "* The student mentioned either one of the following in the explanation:
            ** No net external non-conservative work is being done, so mechanical energy is conserved for the system.
            ** the slide is frictionless, or that the slide is smooth, so that mechanical energy is conserved. Note that the explanation must have indicated using mechanical energy/enery/work principles.
            ** gravity is the only force that does work on the girl.
            ** No non-conservative forces do work on the system."
# Item 3:
     * The student explanation indicated either one of the following:
            ** potential energy or gravitational potential energy is converted or turned into kinetic energy.
            ** Work done by gravity or gravitational force is equal to the change in kinetic energy of the girl or the swimmer.
            ** Discussed relation between work done by gravity and kinetic energy or the velocity of the girl/swimmer.
            ** Discussed the relation between potential energy or height, and the final kinetic energy (or the girl's velocity), when the rest of the explanation resolves around energy concepts.
      * The student can express potential energy as mgh or mgy, and kinetic energy as 1/2mv^2 or 0.5 mv^2. They can write expression such as mgh = 1/2mv^2"
```

Question 2:
**Simple Rubric:**
```
# Item 1: The student wrote down conservation of mechanical energy equation or indicated that mechanical energy can be used to solve the problem

# Item 2: The potential energy term of the conservation of mechanical energy formula contains both a gravitational potential energy term and an elastic potential energy term.

# Item 3: The gravitational potential energy term contains an expression similar to mg(h + L - L_0), and shouldn't be just mgh or mgL
```

*Contact author: Zhongzhou.Chen@ucf.edu

**Detailed Rubric:**
# Item 1: The student wrote down conservation of mechanical energy equation or indicated that mechanical energy can be used to solve the problem
   * The student could write mathematical expressions such as MEi = MEf, ME_i – ME_f = 0, or KE_i + PE_i = KE_f + PE_f, or other similar forms
   * Students could use terms such as "Energy", or "Mechanical Energy".

# Item 2: The potential energy term of the conservation of mechanical energy formula contains both a gravitational potential energy term and an elastic potential energy term.
   * The student must mention both gravitational potential energy and elastic potential energy in the solution, mentioning only one of the two will not satisfy this rubric item.
   * Elastic potential energy could be implied in mathematical expressions such as 1/2k(L–L0)^2, or 0.5*k*x^2, or 0.5*k(L0-L)^2.
   * For this rubric item only, gravitational potential energy could be implied in mathematical forms consisting of mg multiplied by a height or distance measure, such as mgh, mg(L–L0), or mg(h-L-L0)

# Item 3: The gravitational potential energy term contains an expression similar to mg(h + L – L_0), and shouldn't be just mgh or mgL
   * The gravitational potential energy term could take forms such as mg(h + L0 – L) or m*g*(h-l+l0), or other forms that involves modifications to the height h.
   * The student could also write expressions such as mgh + mg(L–L0) or mgh + mg*(L_0 – L) or similar forms
   * Stating that gravitational potential energy is mgh, or including mgh alone will not satisfy this rubric item.

**Detailed Rubric 2:**
# Item 1: The student wrote down conservation of mechanical energy equation or indicated that mechanical energy can be used to solve the problem
   * The student could write mathematical expressions such as MEi = MEf, ME_i – ME_f = 0, or KE_i + PE_i = KE_f + PE_f, or other similar forms
   * Students could use terms such as "Energy", or "Mechanical Energy".
   * The student could also write the kinetic and potential energy terms, such as 0.5mv^2, 1/2kx^2, mgh, mg(h–L), or similar terms, without explicitly mentioning mechanical energy.

# Item 2: The student solution included both a gravitational potential energy term and an elastic potential energy term.
   * The student must mention both gravitational potential energy and elastic potential energy in the solution, mentioning only one of the two will not satisfy this rubric item.
   * Elastic potential energy could be implied in mathematical expressions such as 1/2k(L–L0)^2, or 0.5*k*x^2, or 0.5*k(L0-L)^2.
   * For this rubric item only, gravitational potential energy could be implied in mathematical forms consisting of mg multiplied by a height or distance measure, such as mgh, mg(L–L0), mg(h-L-L0), mg(h–L0)
   * The final elastic potential energy is zero, so elastic potential energy can be omitted in the final potential energy of the situation.

# Item 3: The student indicated that the calculation of the gravitational potential energy involves a modification to the height h.
   * The gravitational potential energy term could take forms such as mg(h + L0 – L) or m*g*(h-l+l0), mg(h–L), mg(h+L0) or other forms that involves the term mg multiplied by a height that is not just h.
   * The student could also write expressions such as mgh + mg(L–L0) or mgh + mg*(L_0 – L), or include both an mgh term and mg(l-l0) on different sides of the equation.
   * The student could also indicate in words that the height in gravitational potential energy can be calculated by adding the spring compression length to the height of the ball.
   * Stating that gravitational potential energy is just mgh, or including mgh alone in the equation will not satisfy this rubric item.

*Contact author: Zhongzhou.Chen@ucf.edu

Question 3:
**Simple Rubric:**
# Item 1: The student solution decomposed the initial linear momentum of boulder 2 into its x and y components.

# Item 2: The student wrote down conservation of linear momentum equation for both the x and y directions independently.

# Item 3: The student used Pythagorean theorem to find the magnitude of the final velocity.

Detailed Rubric:
# Item 1: The student considered the x and y components of the linear momentum of the second boulder separately.
    * The student could write down the x and y components of linear momentum of the second boulder in two separate conservation of linear momentum equations.
    * The student could decompose the linear momentum of the second boulder using trigonometry, for example m_2v2cos(theta), m2v2sin(theta), or mvcos(theta), mvsin(theta) for the second boulder.
    * The student could explicitly state decomposing linear momentum of the second boulder into its x and y components.
    * The student may also decompose the velocity of the second boulder into its x and y components.

# Item 2: The student wrote down conservation of linear momentum equations for both the x and y directions independently.
    * The student must write down equations for both x and y directions.
    * Conservation of linear momentum equations can take forms such as m1v1x + m2v2x = (m1 + m2)vx, and m2v2y = (m1 + m2)vy.
    * The student could imply that linear momentum is conserved on both x and y directions, or conservation of linear momentum procedure is applied to both directions.
    * Only stating that linear momentum is conserved or writing down a single equation such as m1v1 + m2v2 = (m1 + m2)V do not satisfy this rubric.

# Item 3: The student used Pythagorean theorem to find the magnitude of the final velocity.
    * The student could write equations such as v_2f^2 = v_2xf^2 + v_2yf^2, or v2f = sqrt(v2xf^2 + v2yf^2)
    * The student could also state that the final velocity is obtained using the pythagorean theorem, or by taking the square root of the velocity squares.
    * The student could also state that the final velocity is the vector sum of the x component and y component velocities.
    * The student could also apply pythagorean theorem directly to the linear momentums, and divide the final momentum by the mass of the boulders.
    * The student could write equations such as (m2v2)^2 = (m2v2x)^2 + (m2v2y)^2

**Detailed Rubric 2:**
# Item 1: The student considered the x and y components of the linear momentum of the second boulder separately.
    * The student could write down the x and y components of linear momentum of the second boulder in two separate conservation of linear momentum equations.
    * The student could decompose the linear momentum of the second boulder using trigonometry, for example m_2v2cos(theta), m2v2sin(theta), or mvcos(theta), mvsin(theta) for the second boulder.
    * The student could explicitly state decomposing linear momentum of the second boulder into its x and y components.
    * The student could also decompose the velocity of the second boulder into its x and y components.

# Item 2: The student applied conservation of linear momentum to both the x and y directions independently.
    * Conservation of linear momentum equations can take forms such as m1v1x + m2v2x = (m1 + m2)vx, and m2v2y = (m1 + m2)vy.
    * The student could imply that linear momentum is conserved on both x and y directions, or conservation of linear momentum procedure is applied to both directions.

*Contact author: Zhongzhou.Chen@ucf.edu

* Only stating that linear momentum is conserved or writing down a single equation such as m1v1 + m2v2 = (m1 + m2)V do not satisfy this rubric.
    * The student could write down one conservation of linear momentum equation, and either immediately or later indicate that this equation is applied to both x and y directions.

# Item 3: The student used Pythagorean theorem to find the magnitude of the final velocity.
    * The student could write equations such as v_2f^2 = v_2xf^2 + v_2yf^2, or v2f = sqrt(v2xf^2 + v2yf^2)
    * The student could also state that the final velocity is obtained using the pythagorean theorem, or by taking the square root of the velocity squares.
    * The student could also state that the final velocity is the vector sum of the x component and y component velocities.
    * The student could write equations such as (m2v2)^2 = (m2v2x)^2 + (m2v2y)^2
    * The student could also apply pythagorean theorem directly to the linear momentums, and divide the final momentum by the mass of the boulders.
    * The student cannot apply pythagorean theorem directly to p1 and p2, or the linear momentum of boulders 1 and 2. The pythagorean theorem must by applied to the components of velocity or momentum.
    * Simply stating "obtain the magnitude" of the velocity do not satisfy this rubric.

**Detailed Rubric 3:**
# Item 1: The student solution decomposed the initial linear momentum of boulder 2 into its x and y components.
    * The student could also write m2v2x and m2v2y in conservation of linear momentum equations without explicitly decomposing the momentum or the velocities.

# Item 2: The student wrote down conservation of linear momentum equation for both the x and y directions independently.
    * The student could also imply that linear momentum is conserved on both x and y directions separately, and may not explicitly write down conservation equations.
    * Only saying "momentum equation" or "used momentum" does not satisfy this rubric.

# Item 3: The student used Pythagorean theorem to find the magnitude of the final velocity.
    * Just saying "put them together" does not satisfy this rubric

# Appendix B: Mean and Range of Accuracy Metrics for Self-Consistency Runs

**Question 1:**

| Metric | mean | range |
|---|---|---|
| SMD.h1 | 0.094 | 0.08 ~ 0.104 |
| match.h1 | 75.8% | 71.9% ~ 79.2% |
| SMD.h2 | 0.097 | 0.09 ~ 0.104 |
| match.h2 | 74.2% | 72.9% ~ 76.0% |
| diff.both | 14.6% | 13.5% ~ 16.7% |
| Macro F1 | 0.917 | 0.91 ~ 0.921 |
| QWK | 0.9 | 0.885 ~ 0.915 |

**Question 2:**
Self-consistency:

| Metric | mean | range |
|---|---|---|
| SMD.h1 | 0.124 | 0.12 ~ 0.14 |
| match.h1 | 66% | 63% ~ 69% |
| SMD.h2 | 0.13 | 0.12 ~ 0.14 |
| match.h2 | 68% | 64% ~ 71% |
| diff.both | 21.8% | 18.0% ~ 26.0% |
| QWK(avg) | 0.752 | 0.73 ~ 0.77 |
| Macro F1 | 0.818 | 0.81 ~ 0.82 |

Self-consistency 2:

| Metric | mean | range |
|---|---|---|
| SMD.h1 | 0.074 | 0.07 ~ 0.09 |
| match.h1 | 79.8% | 76.0% ~ 81.0% |
| SMD.h2 | 0.088 | 0.08 ~ 0.1 |
| match.h2 | 76.6% | 75.0% ~ 78.0% |
| diff.both | 10.2% | 8.0% ~ 14.0% |
| QWK(avg) | 0.878 | 0.87 ~ 0.89 |
| Macro F1 | 0.864 | 0.85 ~ 0.87 |

**Question 3:**
Self-consistency:

| Metric | mean | range |
|---|---|---|
| SMD.h1 | 0.168 | 0.14 ~ 0.19 |
| match.h1 | 62.6% | 56.0% ~ 68.0% |
| SMD.h2 | 0.124 | 0.1 ~ 0.14 |
| match.h2 | 68.2% | 65.0% ~ 74.0% |
| diff.both | 27% | 20% ~ 32% |
| QWK(avg) | 0.852 | 0.82 ~ 0.88 |
| Macro F1 | 0.846 | 0.83 ~ 0.87 |

Self-consistency 3:

| Metric | mean | range |
|---|---|---|
| SMD.h1 | 0.156 | 0.15 ~ 0.17 |
| match.h1 | 63.8% | 60.0% ~ 66.0% |
| SMD.h2 | 0.1 | 0.09 ~ 0.11 |
| match.h2 | 74.6% | 70.0% ~ 78.0% |
| diff.both | 19.6% | 17.0% ~ 22.0% |
| QWK(avg) | 0.856 | 0.85 ~ 0.87 |
| Macro F1 | 0.846 | 0.84 ~ 0.85 |

*Contact author: Zhongzhou.Chen@ucf.edu

## Appendix C: Prompt for feedback generation.

System message:
```
You are a college introductory level physics teacher who writes kind and concise feedback to
students based on the following information: the problem body, the student's answer to the problem,
the grading rubric, and the teacher's grading justification.
```
Human message template:
```
    This is the problem that the student solved:
    {ProblemBody}
    The student wrote this answer explaining their problem solving process:
    {StudentAnswer}
    The answer was graded according to this grading rubric:
    {Rubric}
    You graded the students' answer and wrote the following grading justification:
    {GradingText}
    Your will write a feedback message to the student following this format:
    {Feedback_format}
```
Feedback Instructions:
```
* First, briefly summarize the the grading standards, by summarizing the grading rubrics.
      ** Start with "Partial credit grading for this problem is based on the following aspects:".
Then summarize each rubric, for example:
            "First, whether you ...."
            "Second, whether your explanation indicated that you...."
            "Finally, whether your explanation showed that ...."
   * Second, point out what the student did well and what the student failed to demonstrate in the
solution, based on students' answer and the grading justification.
         ** If a student satisfies one or more rubric items (i.e. receiving a grade 1), acknowledge
that by saying something similar to the following:
            *** "You clearly showed that...."
            *** "You did a good job explaining that ..."
            *** "You mentioned in your explanation that... which I interpreted as ....."
            Skip this part if the student solution didn't satisfy any of the rubric items.
         ** If a student missed one or more rubric items (i.e. receiving a grade 0), point it out by
starting the sentence similar to the following:
            *** "However, you did not.... "
            *** "However, your explanation did not show that you ...."
            *** "However, I don't think that you ...."
         ** If a student didn't satisfy any of the grading rubrics (i.e. the grade is 0,0,0), say
for example:
            *** "Unfortunately, your explanation didn't indicate any of the aspects above."
            *** "Unfortunately, I don't think your explanation mentioned any of the aspects listed
above."
   * Finally, tell the student "if you have any questions regarding the grading, do not hesitate
to reach out to your teacher."
   * Limit the feedback to this problem, and do not give any study suggestions such as read a
certain book chapter.
   * Also, do not mention words such as "rubric items", since students have not seen the rubric.
```

*Contact author: Zhongzhou.Chen@ucf.edu

**Appendix D: Instructions provided to students during the exam**

**Question 1:**

Please explain:
1. Which physics principle, if any, did you use in your reasoning?
2. Why do you think the physics principle can be applied in this situation.
3. Explain the steps of your reasoning on how you reached your conclusion.

**Question 2 and Question 3:**

Please explain the approach that you took to solve the problem above. Your explanation will be used by the course graders to assign partial credit to you after the exam. (You will not upload a written solution after the exam).
**Note: The instructions for this problem is slightly different from previous ones.**
To be considered for partial credit:
  1. Your explanation CANNOT contain any numbers, but can contain math symbols such as F_x, N, k. Please do not use the equation editor since it messes up the output.
Your explanation should explicitly contain all of the following information:
  1. What variable was the problem asking for and what are the given information? (For example, I'm solving for the final y component velocity of a projectile motion, given the initial height h, velocity v, and launch angle theta). This ensures that the TA understands your notations.
  2. For this problem, explain how many stages did you break the process into, and which conservation law is applicable in each stage and why.
   4. For each step, explain:
      a. What physics principle or equation did you use? (For example, Newton's first/second law of motion, or definition of kinetic energy)
      b. What quantity did you find and how? (For example, I found the normal force by adding all forces on the y direction)
      c. What other key information did you consider (For example, since y-coordinate system is pointing up, gravity is negative)
Note: Some steps, such as setting up coordinate system or drawing free body diagrams may not have an explicit outcome
Note: You can write out formulas such as N + (-mg) = 0. Please do not use the equation editor since it messes up the output.
Note: If you are stuck or not sure how to solve this problem, do not write lengthy explanations on why you are stuck.

*Contact author: Zhongzhou.Chen@ucf.edu